\documentclass[12pt,twoside]{article}

\usepackage{jeep}       
\mysection{section}{\large \bf}{\thesection.~}   
\mysection{subsection}{\normalsize\bf}{\thesubsection.~}

\usepackage{a4wide}     
\usepackage{fancyheadings}

\usepackage{cite}   
\usepackage{epsf}

\advance \headheight by 3.0truept       

\usepackage{times}

\newcommand{\lt}{\left}
\newcommand{\rt}{\right}
\newcommand{\ov}{\overline}

\newcommand{\eq}[1]{(\ref{#1})}

\def\openone{\leavevmode\hbox{\small1\kern-3.8pt\normalsize1}}%

\newcommand{\no}{\nonumber }

\newcommand{\imag}{{\rm Im}\,}
\newcommand{\real}{{\rm Re}\,}

\newcommand{\prl}{Phys.~Rev. Lett.}
\newcommand{\prd}{Phys.~Rev.~D}
\newcommand{\plb}{Phys.~Lett.~B}
\newcommand{\npb}{Nucl.~Phys.~B}

\newlength{\miniwidth}
\newlength{\miniwidthplot}
\newlength{\nseparation}
\newenvironment{nfigure}[1]
        {\begin{figure}[#1]\hrule\vspace{\nseparation}\par}
        {\vspace{\nseparation}\par \hrule \end{figure}}

\begin{document}
~\\[-20truemm]
KEK-TH-616, FERMILAB-Pub-99/035-T, DPNU-99-10 \hfill 
\parbox{4cm}{hep-ph/9903230}\\
~\vspace{0.5truecm}
\begin{center}
{\LARGE\bf 
A short look at $\mathbf{\epsilon^\prime/\epsilon}$}\\[2\baselineskip]
\textsl{
Yong-Yeon Keum\footnote{e-mail:keum@ccthmail.kek.jp, Monbusho fellow},\\
KEK, Theory Group, 
Tsukuba, Ibaraki 305-0801, Japan,  \\[2mm] 
Ulrich Nierste\footnote{e-mail:nierste@fnal.gov},\\
Fermilab, Theory Division MS106, IL-60510-500, USA,\\[2mm]
and \\[2mm] 
A.I.~Sanda\footnote{e-mail:sanda@eken.phys.nagoya-u.ac.jp} \\
Dept.\ of Physics, Nagoya University, Nagoya 446, Japan}  \\
\end{center}
%
\vspace{5mm}
\begin{center}
\textbf{\large Abstract} 
\end{center}
We analyze the theoretical implications of the new KTeV measurement of direct
CP-violation in $K\rightarrow
\pi \pi$ decays. The result 
is found consistent with the Standard Model for low values of the
strange quark mass $m_s$. If the hadronic parameters $B_6^{(1/2)}$ and
$B_8^{(3/2)}$ satisfy $2 B_6^{(1/2)} - B_8^{(3/2)}\leq 2$, as suggested
by lattice and $1/N_c$ calculations, we find an upper bound of $ 110
\, \mathrm{MeV}$ for $m_s(2 \, \mathrm{GeV})$.  We parametrize
potential new physics contributions to $\epsilon^\prime/\epsilon$ and
illustrate their correlation with upper bounds on $m_s$.  Finally we
discuss a non-perturbative mechanism, which is not contained in the
existing calculations of $B_6^{(1/2)}$. This mechanism  enhances
$B_6^{(1/2)}$ and thereby leads to a better understanding of the
$\Delta I=1/2$ rule and the high measured value of $\real
(\epsilon^\prime/\epsilon)$.\\[2mm]
{\small PACS: 11.30.Er; 13.25.Es; 14.65.Bt}
%
%
%
\begin{center}
-------------------------------------------------------------------
\end{center}
Recently the KTeV collaboration at Fermilab has precisely determined
the measure of direct CP-violation in $K\rightarrow \pi \pi$ decays
\cite{KTeV}:
\begin{eqnarray}
\real (\epsilon^\prime/\epsilon)&=&(28\pm4)\cdot 10^{-4}. \label{eps}
\end{eqnarray}
This measurement is consistent with the result of the CERN experiment
NA31, which has also found a non-vanishing value for $\real
(\epsilon^\prime/\epsilon)$ \cite{na}.  Within the last two decades a
tremendous effort has been made to calculate the short distance QCD
effects with next-to-leading order accuracy \cite{bur} and to obtain
the relevant hadronic matrix elements using various non-perturbative
methods
\cite{latt,nc,chqm}. Yet while the Standard Model predicts 
$\epsilon^\prime/\epsilon$ to be non-zero, the theoretical prediction
of its precise value is plagued by large uncertainties due to an
unfortunate cancellation between two hadronic quantities. Nevertheless
for the ballpark of popular input parameters $ \real
(\epsilon^\prime/\epsilon)$ results in $2$--$16$ times $10^{-4}$
\cite{rev1}, so that the large value in \eq{eps} came as a
surprise to many experts. Here a key role is played by the strange
quark mass, whose size is not precisely known at present. In the
Standard Model $\real (\epsilon^\prime/\epsilon)$ can be summarized in
the handy formula \cite{rev1,bs}:
\begin{eqnarray}
\real (\epsilon^\prime/\epsilon)&=& 
 \imag \lambda_t \, \lt[ -1.35 + R_s \lt( 1.1 \lt| r_Z^{(8)} \rt|
   B_6^{(1/2)} + \lt( 1.0 -0.67 \lt| r_Z^{(8)} \rt| \rt) 
	B_8^{(3/2)}  \rt)\rt]. 
\label{sm}
\end{eqnarray}
Here $\lambda_t = V_{td} V_{ts}^*$ is the CKM factor and $r_Z^{(8)}$
comprises the short distance physics. Including  
next-to-leading order QCD corrections the short distance factor is in
the range  \cite{bur,rev1}
\begin{eqnarray}
 6.5 \leq \lt| r_Z^{(8)} \rt| \leq 8.5 \label{rz}
\end{eqnarray}   
for $0.113 \leq \alpha_s^{\ov{\rm MS}} (M_Z) \leq 0.123$. Relevant
contributions to $\epsilon^\prime/\epsilon$ stem from the $\Delta
I=1/2$ matrix element of the operator $Q_6$ and the $\Delta I=3/2$
matrix element of $Q_8$ (see \cite{bur,rev1,rev2} for their precise
definition).  These hadronic matrix elements are parametrized by
$B_6^{(1/2)}$ and $B_8^{(3/2)}$. Finally the dependence on the strange
quark mass is comprised in
\begin{eqnarray}
R_s &=& \lt( \frac{150 \, \mathrm{MeV}}{m_s \lt( m_c \rt)}   \rt)^2. \no
\end{eqnarray}
From standard analyses of the unitarity triangle \cite{hn,bur2} one
finds 
\begin{eqnarray} 
1.0 \cdot 10^{-4} \leq \imag \lambda_t 
	\leq 1.7 \cdot 10^{-4} .\label{lat}
\end{eqnarray} 
Lattice calculations \cite{latt} and the $1/N_c$ expansion \cite{nc} predict 
\begin{eqnarray}
0.8 \leq B_6^{(1/2)} \leq 1.3, \qquad
0.6 \leq B_8^{(3/2)} \leq 1.0.
\label{Brange}
\end{eqnarray}
The maximal possible $\real
(\epsilon^\prime/\epsilon)$ for the quoted ranges of the input parameters
is plotted vs.\ $m_s$ in Fig.~\ref{fig1}.
Hence if the Standard Model is the only source of
direct CP-violation in $K\rightarrow \pi \pi$ decays, 
the 2$\sigma$ bound from \eq{eps}, 
$\real (\epsilon^\prime/\epsilon) \geq 20 \cdot 10^{-4}$, implies 
\begin{eqnarray} 
m_s (m_c) \leq 126 \, \mathrm{MeV} \quad &\Leftrightarrow& \quad
m_s \lt(2 \, \mathrm{GeV}  \rt) \leq 110 \, \mathrm{MeV}. 
\label{ms}
\end{eqnarray} 
in the $\ov{\rm MS}$ scheme. The upper bounds in \eq{ms} correspond to
the maximal values for $\imag \lambda_t$, $2\, B_6^{(1/2)} -
B_8^{(3/2)}$ and $\lt| r_Z^{(8)} \rt|$.  In the chiral quark model
\cite{chqm,rev2} $B_6^{(1/2)}$ can exceed the range in \eq{Brange} and
$2\, B_6^{(1/2)} - B_8^{(3/2)}$ can be as large as 2.9 relaxing the
bound in \eq{ms} to $m_s (m_c) \leq 151 \, \mathrm{MeV} $.  In
\cite{rev2} it has been argued that this feature of the chiral quark
model prediction should also be present in other approaches, once
certain effects (final state interactions, ${\cal O}(p^2)$ corrections
to the electromagnetic terms in the chiral lagrangian) are
consistently included.  Hence the result in \eq{eps} is perfectly
consistent with values for $m_s$ obtained in quenched lattice
calculations favouring $m_s \lt(2 \, \mathrm{GeV} \rt) = \lt( 110 \pm
30 \rt)$ MeV \cite{lattmass}.  From unquenched calculations one
expects even smaller values \cite{bh}.  It is also consistent with
recent sum rule estimates \cite{summass}. However the preliminary
ALEPH result for the determination of $m_s$ from $\tau$ decays,
$m_s(m_{\tau}) =(172\pm 31)$ MeV \cite{aleph}, violates the bound in
\eq{ms}. The compatibility of the ranges in \eq{lat} and \eq{Brange}
with large values of order ${\cal O} (2\cdot 10^{-3})$ for $\real
(\epsilon^\prime/\epsilon) $ has been pointed out earlier in
\cite{burbig}. Here instead we aim at the most conservative upper
bound on $m_s$ from \eq{rz}, \eq{lat}, \eq{Brange} and the
experimental result in \eq{eps}, as quoted in \eq{ms}.

With the present uncertainty in \eq{Brange} and in the lattice
calculations of $m_s$ one cannot improve the range for $\imag
\lambda_t$ in \eq{lat}. Hence at present $\epsilon^\prime/\epsilon$
is not useful for the construction of the unitarity triangle.

While we do not claim the necessity for new physics in
$\epsilon^\prime/\epsilon$, there is certainly plenty of room for it in
$\epsilon^\prime/\epsilon$ and other observables in the Kaon system
such as $\epsilon_K$ or $\Delta M_K$ \cite{hn2} or rare K decays
\cite{buc}. Now
\eq{eps} correlates non-standard contributions to $\epsilon^\prime/\epsilon$
with upper bounds on $m_s$, which might become weaker or stronger
compared to \eq{ms} depending on the sign of the new physics
contribution. We want to stress that this feature is very useful to
constrain new physics effects in other $s\rightarrow d$ transitions:
Most extension of the Standard Model involve new helicity-flipping
operators, for example $\epsilon_K$ can receive contributions from
the $\Delta S=2$ operator $Q_S=\ov{s}_L d_R \, \ov{s}_L d_R$, which is
absent in the Standard Model. Yet the matrix elements of operators
like $Q_S$, which involves two (pseudo-)scalar couplings, are
proportional to $1/m_s^2$. Hence upper bounds on $m_s$ imply 
\textit{lower} bounds on the matrix elements of $Q_S$ and similar
operators multiplying the new physics contributions of interest. 
To exploit this feature one must, of course, first explore the
potential impact of the considered new model on 
$\epsilon^\prime/\epsilon$. Recently Buras and Silvestrini \cite{bs}
have pointed out that $\epsilon^\prime/\epsilon$ is sensitive to new 
physics contributions in the effective $\ov{s}dZ$-vertex.  
This vertex can be substantially enhanced in generic supersymmetric
models, as discovered by Colangelo and Isidori \cite{ci}. By contrast 
supersymmetric contributions to the gluonic penguins entering 
$\epsilon^\prime/\epsilon$ are small \cite{gms}. We want to
parametrize the new physics in a model independent way and write 
\begin{eqnarray} 
\!\!
\lt. \real (\epsilon^\prime/\epsilon) \rt|_{new} \!\!\!\!&=&\!\!  
   \imag Z_{ds}^{new} \lt[ 1.2 - R_S |r_Z^{(8)}| B_8^{(3/2)} \rt] +
   \imag C_{ds}^{new} \cdot 0.24 + 
   15 \cdot 10^{-4} R_s \, B_6^{(1/2)} R_6 \label{np}  	
\end{eqnarray}
Here $Z_{ds} ^{new}$ is the new physics contribution to the effective
$\ov{s}dZ$-vertex $Z_{ds}$ defined in \cite{bs}. $C_{ds}$ is the effective 
chromomagnetic $\ov{s}dg$-vertex defined by 
\begin{eqnarray} 
{\cal L} &=& \frac{G_F}{\sqrt{2}} C_{ds} \cdot Q_{11} \lt( M_W \rt), 
\qquad 
Q_{11} \; = \; \frac{g_s}{16 \pi^2} m_s \ov{s} \sigma^{\mu \nu} T^a
		(1-\gamma_5) d \, G_{\mu \nu}^a \label{chr}.
\end{eqnarray} 
The impact of the chromomagnetic operator $Q_{11}$ has been analyzed
in \cite{bef}. In the Standard Model one has $C_{ds}=-0.19 \lambda_t$
with negligible impact on $\epsilon^\prime/\epsilon$. In extensions of
the Standard Model, however, $C_{ds}$ can be larger by an order of
magnitude or more, because the factor of $m_s$ in \eq{chr}
accompanying the helicity flip of the $s$-quark may be replaced by the
mass $M$ of some new heavy particle appearing in the one-loop
$\ov{s}dg$-vertex \cite{bk,bk2}. There are no constraints on $\imag
C_{ds}^{new}$ from $\epsilon_K$ or $\Delta m_K$. In the $\ov{b}sg$-vertex
the corresponding enhancement factor is smaller by a factor of
$m_s/m_b$.  The numerical factor of $0.24$ in \eq{np} incorporates the
renormalization group evolution from $M_W$ down to 1 GeV and the
hadronic matrix element calculated in \cite{bef}. Finally new physics
could enter the Wilson coefficient $y_6 (1\, \mathrm{GeV}) \approx
-0.1$ multiplying $B_6^{(1/2)}$ (and hidden in $|r_Z^{(8)}|$ in \eq{sm}).  
(For definitions and numerical values of the Wilson coefficients see
\cite{rev1,rev2,bur2}.) The parameter $R_6$ in \eq{np} is defined as 
\begin{eqnarray} 
R_6 &=& \frac{ \imag \lt[ \lambda_t \, y_6^{new} ( 1 \mathrm{GeV})
		\rt]}{ - 0.17 \cdot 10^{-4} } . \label{r6}
\end{eqnarray} 
Hence $R_6=1$ means that the new physics contribution to $y_6 ( 1
\mathrm{GeV})$ is approximately equal to the Standard Model contribution.
There is no simple relation between $y_6 (1 \mathrm{GeV})$ and the new
physics amplitude at $\mu=M_W$, because the initial values of all QCD
operators contribute to $y_6 ( 1 \mathrm{GeV})$ due to
operator mixing. In a given model one has to calculate these initial
coefficients and to perform the renormalization group evolution down
to $\mu= 1\, \mathrm{GeV}$. In $R_6$ no order-of-magnitude enhancement
like in $C_{ds}$ is possible. Only small effects have been found in 
\cite{rev2}, because $y_6(1\, \mathrm{GeV})$ is largely an admixture of
the tree-level coefficient $y_2(M_W)$, which is unaffected by new physics.
While still  $R_6$ can be more
important than $C_{ds}$ due to the larger coefficient in \eq{np}, it
will be less relevant than $Z_{ds}^{new}$.  In
Fig.\ \ref{fig2} we have plotted the correlation between 
$\imag Z_{ds}^{new}$ and $m_s$ for $C_{ds}=R_6=0$. We have used the range in
\eq{lat}. The upper bound on $-\imag Z_{ds}^{new}$ is related to the
lower bound on $\imag \lambda_t$, which can be invalidated, if new
physics contributes to $\epsilon_K$. The more interesting lower
bound, however, corresponds to the upper limit in \eq{lat} stemming
from tree-level semileptonic $B$ decays, which are insensitive to new
physics. 

Maybe future determinations of $m_s$ and more precise measurements of
$\real (\epsilon^\prime/\epsilon)$ will eventually be in conflict with
\eq{ms}.  Before then discussing the possibility of new physics it is
worthwile to consider, if $B_6^{(1/2)}$ can be increased over the
maximal value quoted in \eq{Brange} by some strong interaction
dynamics.  In \cite{sanda} it has been pointed out that the existence
of $f_0(400-1200)$, a $\pi\pi$ S-wave $I=0$ resonance, introduces a
pole in the $\Delta I=1/2$ matrix element of $Q_6$.  This mechanism is
not contained in standard chiral perturbation theory and therefore not
included in the calculations leading to \eq{Brange}. It can lead to a
factor of 2-4 enhancement of $B_6^{(1/2)}$ allowing to relax the upper
limit in \eq{ms}.  $B_6^{(1/2)}$ also enters the real part of $\Delta
I=1/2$ amplitude $A_0$, whose large size is an yet unexplained puzzle
of low energy strong dynamics ($\Delta I=1/2$ rule).\footnote{We
consider an explanation in terms of a new physics enhancement of the
chromomagnetic vertex $C_{ds}$ as proposed in \cite{bk2} to be
unlikely in view of the small matrix element found in \cite{bef}.}
Now with the large measured value for $\real
(\epsilon^\prime/\epsilon)$ an enhancement of $B_6^{(1/2)}$ becomes
phenomenologically viable.  We have extracted the maximum value of
$B_6^{(1/2)}$ compatible with \eq{eps}, \eq{rz} and \eq{lat}. The
result is plotted vs.\ $m_s$ in Fig.~\ref{fig3}. Subsequently we have
inserted the extracted result for $(B_6^{(1/2)},m_s)$ together with
the $1/N_c$ predictions for $B_1^{(1/2)}$ and $B_2^{(1/2)}$
\cite{nc,rev1} into the theoretical prediction for $\real A_0$ and
solved for the Wilson coefficient $z_6 (\mu \approx 1\,
\mathrm{GeV})$.  $B_6^{(1/2)}$ depends only weakly on $\mu$
\cite{bur,rev1,rev2}.  The numerical value of $z_6 (\mu) $ suffers
from severe scheme and scale dependences \cite{bur,rev1,bur2}. We
could fit the measured result for $\real A_0$ with a value for $z_6
(\mu)$, which exceeds the value of $z_6 (1\, \mathrm{GeV})$ in the NDR
scheme by less than a factor of 2. The extracted value depends only
weakly on $m_s$ and slightly decreases with increasing $m_s (m_c)$.
Considering the large uncertainty in $z_6 (1\, \mathrm{GeV})$ and the
fact that the true scale of the hadronic interaction is probably well
below $1\, \mathrm{GeV}$ ($|z_6|$ increases with decreasing $\mu$, but
cannot reliably be predicted for too low scales) we conclude that the
mechanism proposed in \cite{sanda} leads to a semiquantitative
understanding of the $\Delta I=1/2$ rule while simultaneously being
consistent with the measurement of $\real (\epsilon^\prime/\epsilon)$
in \eq{eps}. This conclusion would not have been possible with the
old low result of the E731 experiment \cite{e731}. Also the upper
bound on $m_s$ in \eq{ms} becomes invalid, so that one can even
accomodate for the high value of $m_s$ measured by ALEPH \cite{aleph}
and quoted after \eq{ms}.

\begin{nfigure}{tb}
\centerline{\epsfxsize=\textwidth \epsffile{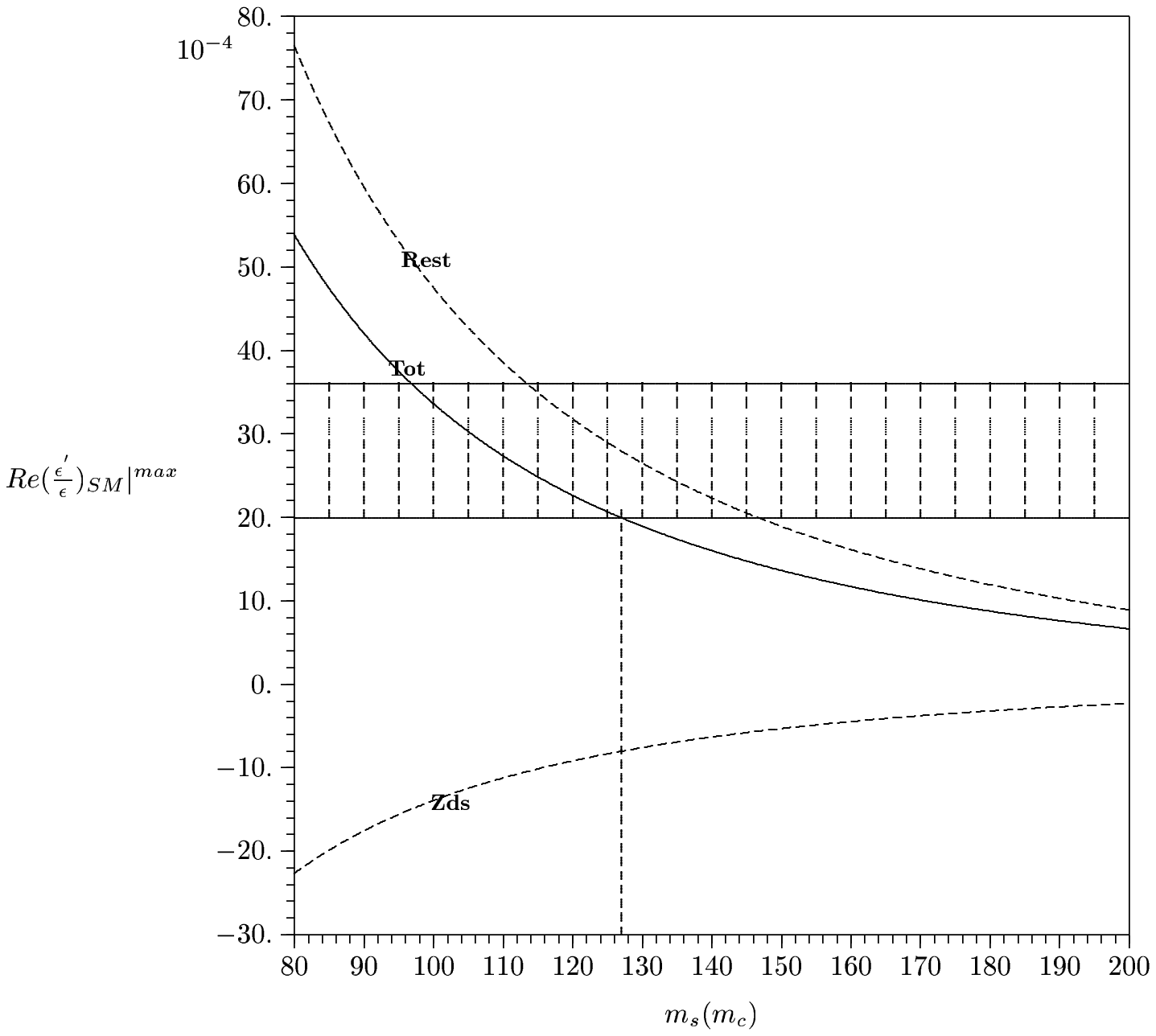}}
\caption{The maximal  
$\real (\epsilon^\prime/\epsilon)$ vs.\ $m_s(m_c)$.  $\lt. \real
(\epsilon^\prime/\epsilon)\rt|_{max}$ corresponds to $\imag
\lambda_t=1.7 \cdot 10^{-4}$, $2\, B_6^{(1/2)} - B_8^{(3/2)}=2.0$ and
$\lt| r_Z^{(8)} \rt| = 8.5$. The plotted relation for different values
of $2\, B_6^{(1/2)} - B_8^{(3/2)}$ can be obtained by replacing $m_s
(m_c)$ with $m_s (m_c) \cdot \lt[ \lt(2\, B_6^{(1/2)} -
B_8^{(3/2)}\rt)/2.0\rt]^{-1/2} $.  The contribution of the
$\ov{s}dZ$-vertex is shown separately.  The hatched area corresponds
to the 2$\sigma$ range of the measured value in \eq{eps}. }\label{fig1}
\end{nfigure}

\begin{nfigure}{tb}
\centerline{\epsfxsize=\textwidth \epsffile{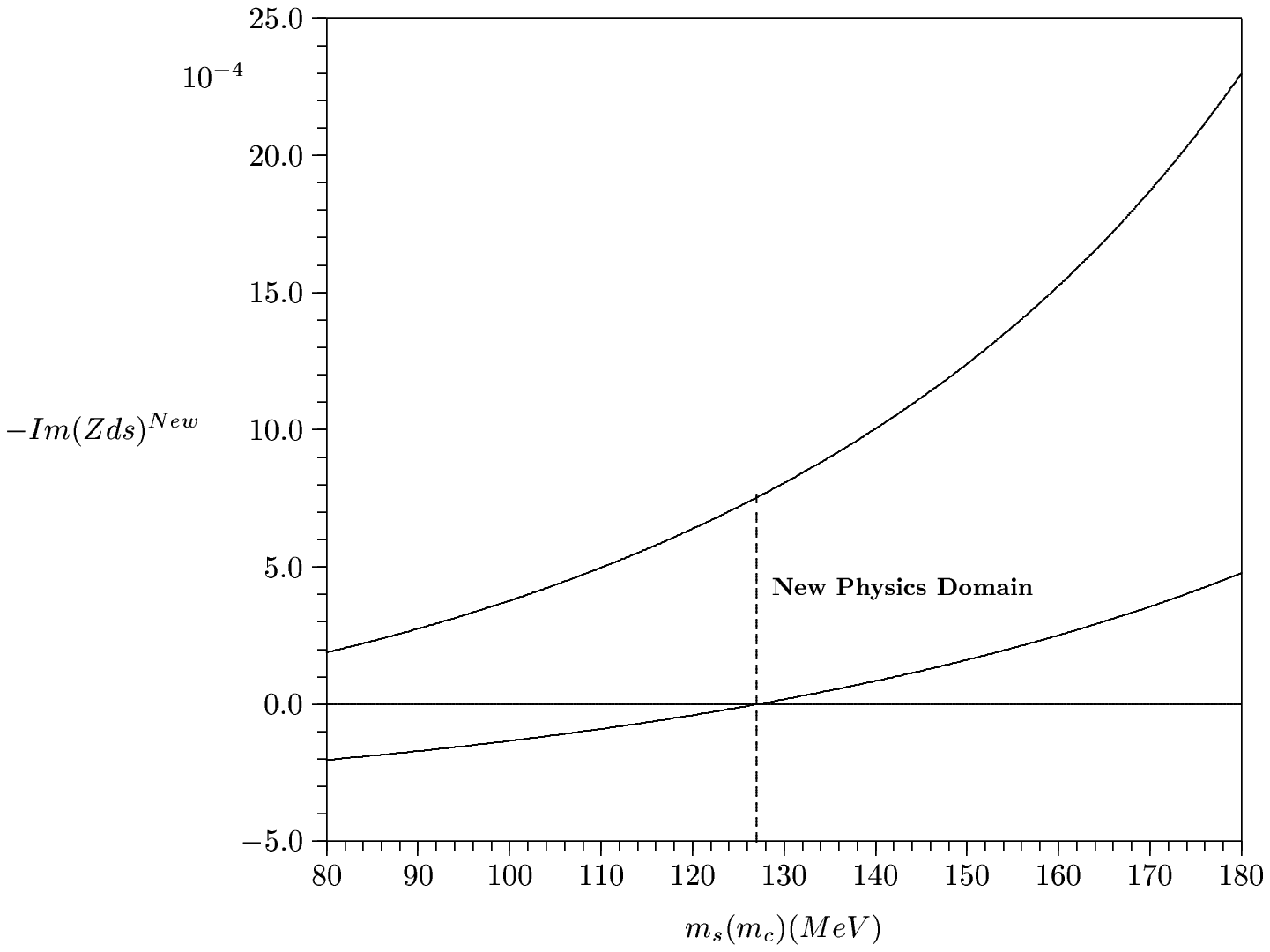}}
\caption{Correlation between new physics contributions to $\imag
Z_{ds}$ and $m_s$ for the ranges in \eq{rz}, \eq{lat} and \eq{Brange}.
To relax the range in \eq{Brange} for the hadronic B-factors see the
caption of Fig.~\ref{fig1}.   }\label{fig2}
\end{nfigure}

\begin{nfigure}{tb}
\centerline{\epsfxsize=0.9 \textwidth \epsffile{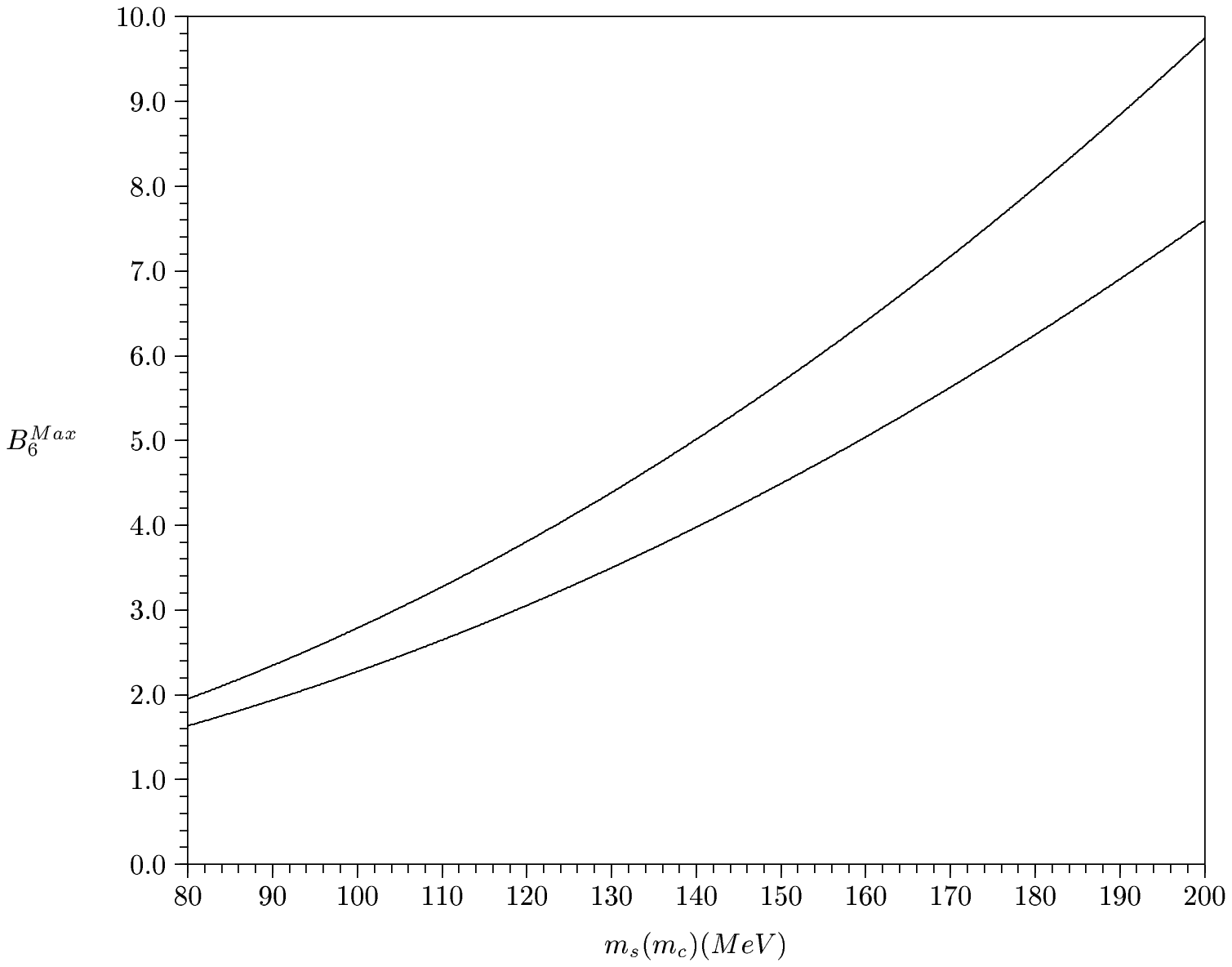}}
\caption{Maximal value of $B_6^{(1/2)}$ vs.\ $m_s$ extracted from
\eq{eps} for $\imag \lambda_t$ and $B_8^{(3/2)}$
in the ranges in \eq{lat} and \eq{Brange}. The lower (upper) curve corresponds
to $\lt| r_Z^{(8)} \rt|=8.5$ ($\lt| r_Z^{(8)} \rt|=6.5$).
   }\label{fig3}
\end{nfigure}

\vspace{5mm}

Acknowledgments: This work has been supported in part by Grant-in-Aid
for Special Project Research (Physics of CP-violation).  Y-Y.K.\ and
U.N.\ appreciate the kind hospitality of Nagoya University, where this
work has been completed. Y-Y.K.\ thanks M.~Kobayashi for
encouragement. He is supported by the Grant-in-Aid of the Japanese 
Ministry of Education, Science, Sports and Culture.

\end{document}